\documentclass[12pt]{article}

\usepackage{latexsym,amsmath,amssymb,amsthm,amsfonts,graphicx}
\usepackage{natbib}
\usepackage{epsfig,bm,authblk,url}
\usepackage{slashbox}

\usepackage{float} 
\usepackage{booktabs} 
\usepackage{graphicx} 
\usepackage[margin=1cm]{caption} 
\usepackage{mathtools}
\usepackage{algorithmicx}
\usepackage{algorithm}
\usepackage{algpseudocode}

\usepackage[title]{appendix} 
\usepackage{enumitem} 
\floatstyle{plain}
\newfloat{Algorithm}{thp}{lop}
\floatname{Algorithm}{Algorithm}

\usepackage{titling}
\def \thetavec {\text{\boldmath$\theta$}}
\def\betavec {\text{\boldmath$\beta$}}
\def\alphavec {\text{\boldmath$\alpha$}}
\def\zetavec {\text{\boldmath$\zeta$}}
\def\Gammamat {\text{\boldmath$\Gamma$}}
\def \muvec {\text{\boldmath$\mu$}}
\def \zerovec {\text{\boldmath$0$}}
\def \Sigmamat {\text{\boldmath$\Sigma$}}
\def \Deltavec {\text{\boldmath$\Delta$}}
\def \Bvec {\text{\boldmath$B$}}
\def \Imat {\text{\boldmath$I$}}
\def \Dmat {\text{\boldmath$D$}}
\def \Rmat {\text{\boldmath$R$}}
\def \Xmat {\text{\boldmath$X$}}
\def \yvec {\text{\boldmath$y$}}
\def \xvec {\text{\boldmath$x$}}
\def \wvec {\text{\boldmath$w$}}
\def \uvec {\text{\boldmath$u$}}
\def \vvec {\text{\boldmath$v$}}
\def \zvec {\text{\boldmath$z$}}
\def \Yvec {\text{\boldmath$Y$}}
\def \Fvec {\text{\boldmath$F$}}
\def \Amat {\text{\boldmath$A$}}
\def \Bmat {\text{\boldmath$B$}}
\def \Tmat {\text{\boldmath$T$}}
\def \Cmat {\text{\boldmath$C$}}

\begin{document}

\title{Assessment and adjustment of approximate inference  \\
algorithms using the law of total variance}
\date{\empty}

\author{Xuejun Yu,\thanks{\textit{Department of Statistics and Applied Probability, National University of Singapore}} \, David J. Nott,\thanks{Corresponding author:  standj@nus.edu.sg. \textit{Department of Statistics and Applied Probability, National University of Singapore} and \textit{Institute of Operations Research and Analytics, National University of Singapore}.} \, Minh-Ngoc Tran\thanks{\textit{Discipline of Business Analytics, The University of Sydney Business School}, 
\textit{Australian Centre of Excellence for Mathematical and Statistical Frontiers (ACEMS)}}\,\, and Nadja Klein\thanks{\textit{School of Business and Economics, Statistics, Humboldt-Universit\"at zu Berlin.}}}

\maketitle

\vspace{-0.7in}

\begin{abstract}
A common method for assessing validity of Bayesian sampling or approximate inference methods makes use of simulated data replicates for parameters drawn from the prior.  
Under continuity assumptions, quantiles of functions of the simulated parameter values within corresponding
posterior distributions are uniformly distributed.  Checking for uniformity when a posterior density is approximated numerically provides a diagnostic
for algorithm validity.  Furthermore, adjustments to achieve uniformity can improve the quality of approximate inference methods.
A weakness of this general
approach is that it seems difficult to extend beyond scalar functions of interest.  The present article
develops an alternative to quantile-based checking and adjustment methods which is 
inherently multivariate.  The new approach is based on use of the tower property of conditional expectation and the law
of total variance for relating prior and posterior expectations and covariances.  
For adjustment, approximate inferences are modified so that the correct prior to posterior relationships hold.  
We illustrate the method in three examples.  The first uses an auxiliary model in a likelihood-free inference
problem.  The second considers corrections for variational Bayes approximations
in a deep neural network generalized linear mixed model.  
Our final application considers a deep neural network surrogate for approximating
Gaussian process regression predictive inference.  


\end{abstract}

\section{Introduction}\label{sec:Intro}

Researchers are increasingly called upon to analyse large datasets using complex models, 
and this has resulted in approximate inference methods being more widely used, such as
likelihood-free inference \citep{marin+prr12,blum+nps13} and variational inference \citep{Ormerod2010,blei+kj17}.
However, often the accuracy of these methods 
cannot be guaranteed, so there is interest in diagnosing algorithm failures, and
in developing adjustments which improve their performance.  The current paper is a contribution to this area, and considers an approach
that can easily handle multivariate quantities of interest, unlike the existing approaches which are hard to extend
beyond checking and adjustment for univariate quantities.  

A common method for checking Bayesian computational algorithms
is based on posterior quantiles \citep{cook+gr06,gelman17}. To save notation, and following the discussion in \cite{cook+gr06}, consider
the case of a scalar parameter $\theta$ of interest.  
The data will be denoted as $\yvec$.  
Suppose we have a joint Bayesian model for $(\theta,\yvec)$ specified by a density $p(\theta,\yvec)=p(\theta)p(\yvec|\theta)$, where $p(\theta)$
is the prior density and $p(\yvec|\theta)$ is the sampling density.  Consider a simulation $(\theta',\yvec ')$ from $p(\theta,\yvec)$, obtained by drawing $\theta'\sim p(\theta)$,
and then $\yvec '\sim p(\yvec|\theta')$.  Noting that $p(\theta)p(\yvec|\theta)=p(\yvec)p(\theta|\yvec)$, we can equivalently think of $(\theta',\yvec')$ as obtained by sampling
$\yvec '\sim p(\yvec)$, then $\theta'\sim p(\theta|\yvec ')$, so that we can regard $\theta'$ in the joint sample as being an exact draw from $p(\theta|\yvec ')$.  
Under continuity assumptions, transforming a random variable by its distribution function gives a uniform random variable, so writing $F(\theta|\yvec ')$ for the posterior
distribution function of $\theta$ given $\yvec '$, we have that $F(\theta'|\yvec ')\sim U[0,1]$, where $U[0,1]$ is the uniform
density on $[0,1]$.  When Markov chain Monte Carlo (MCMC) simulation or some approximate inference algorithm is used to approximate $F(\theta|\yvec ')$ by $\tilde{F}(\theta|\yvec ')$ say, 
testing for uniformity of $\tilde{F}(\theta'|\yvec ')$ based on replicate samples of $(\theta',\yvec ')$ 
can be used to diagnose problems with the simulation algorithm, or to make
inference adjustments.  

\cite{gelman17} points out that problems can occur in implementation of the method of \cite{cook+gr06} if proper account is not taken of correlation between samples
when checking MCMC sampling algorithms.   Difficulties can also arise from discretization when using sample based estimates of quantiles.  \cite{talts+bsvg18} suggest an alternative approach
based on ranks and ways of adjusting for dependence, if that is an issue.   They also show how to interpret the nature of any problems with the algorithm based on the kind
of departure from uniformity observed.  Closely related methods to those of \cite{cook+gr06} have been applied for checking likelihood-free and variational inference algorithms  \citep{wegmann+le09,prangle+bps14,yao+vsg18,lee+nr19}.  
In the context of likelihood-free inference using approximate Bayesian computation (ABC), these quantile-based checks have motivated recalibration 
algorithms to improve the quality
of posterior approximations \citep{rodrigues+ps17}.  In these approaches, if simple importance sampling ABC computational methods are used, 
then repeated posterior approximations for different data can be done using the same prior samples, and
the recalibration adjustments can be performed efficiently.  \cite{rodrigues+ps17} also note that their adjustments are feasible 
when fast approximations are available, and they discuss in detail calibration of inferences based on auxiliary models.  Variational approximations can also 
benefit from the suggested adjustments.

One difficulty with the quantile-based adjustment methods mentioned above is that it seems difficult to apply them in the case of a multivariate $\theta$
when corrections in the dependence structure of the posterior are desired.  In the next section we outline a novel alternative to the standard quantile-based inference adjustments.  
In particular, we consider applying the tower property of conditional expectation and
law of total variance to relate prior and posterior means and covariances.  When the posterior means and covariances are estimated by an approximate inference
algorithm, we suggest adjustments so that the correct prior to posterior relationships hold after the adjustment.  

After outlining the main idea in Section 2, Section 3 explains
the detailed implementation in the case where particle approximations of the posterior distributions are available.  Section 4 considers several applications involving likelihood-free inference, 
and checking and adjusting for approximate inferences in some deep neural network and Gaussian process regression models.   Section 5 gives some concluding discussion.  

\section{The basic idea}\label{sec:basic}

Suppose that $\thetavec$ is a possibly multivariate unknown of interest.  
We define
a Bayesian model through a joint density $p(\thetavec,\yvec)=p(\thetavec)p(\yvec|\thetavec)=p(\yvec)p(\thetavec|\yvec)$.  We
assume that the first two moments of the prior distribution exist.  The tower property of conditional expectation allows us to write
the prior mean as
\begin{align}
  E(\thetavec)= & E(E(\thetavec|\Yvec)),   \label{priormean}
\end{align}
for $\Yvec \sim p(\yvec)$. Now suppose some approximate inference algorithm is available, approximating the true posterior density $p(\thetavec|\yvec)$ by $\tilde{p}(\thetavec|\yvec)$, with 
approximate posterior mean and posterior covariance matrix $\widetilde{E}(\thetavec|\yvec)$ and $\widetilde{\text{Cov}}(\thetavec|\yvec)$ respectively.  If our approximate inference algorithm 
is accurate, then 
\begin{align}
  E(\thetavec) \approx & E(\widetilde{E}(\thetavec|\Yvec)). \label{apmean}
\end{align}
Similarly, applying the law of total variance to the prior covariance matrix $\text{Cov}(\thetavec)$ of $\thetavec$, 
\begin{align}
  \mbox{Cov}(\thetavec) & = E(\text{Cov}(\thetavec|\Yvec))+\text{Cov}(E(\thetavec|\Yvec)), \label{priorcov}
\end{align}
for $\Yvec\sim p(\yvec)$. Once again, if our approximate inference algorithm is accurate,
\begin{align}
  \mbox{Cov}(\thetavec) & \approx E(\widetilde{\text{Cov}}(\thetavec|\Yvec))+\text{Cov}(\widetilde{E}(\thetavec|\Yvec)), \label{apcov}
\end{align}
for $\Yvec\sim p(\yvec)$.  The main idea of our checking and adjustment method is that given independent samples $(\thetavec^{(i)},\yvec^{(i)})\sim p(\thetavec)p(\yvec|\thetavec)$, $i=1,\dots, I$, 
we can estimate the left-hand 
sides of (\ref{apmean}) and (\ref{apcov}) based on the prior samples $\thetavec^{(i)}$ (if the prior  mean and covariance are not available analytically), 
and the right-hand sides of (\ref{apmean}) and (\ref{apcov}) based on the samples
$\yvec^{(i)}$ and the approximations $\widetilde{E}(\thetavec|\yvec^{(i)})$ and $\widetilde{\text{Cov}}(\thetavec|\yvec^{(i)})$.   
For checking, we can assess whether departures from equality in (\ref{apmean}) and (\ref{apcov}) are of practical
concern.  For inference adjustments, we can alter $\tilde{p}(\thetavec|\yvec^{(i)})$ 
so that exact equality holds in (\ref{apmean}) and (\ref{apcov}) for $\widetilde{E}(\thetavec|\yvec^{(i)})$ and $\widetilde{\text{Cov}}(\thetavec|\yvec^{(i)})$  after adjustment.  
Although this is the basic idea, some further refinements are outlined below.  Checking Bayesian inferences according to 
equation (\ref{apmean}) is somewhat related to the idea of calibration of posterior means discussed in \citet{gelman+csdvr13}, p. 128, 
although the use of (\ref{apmean}) and (\ref{apcov}) jointly to check and adjust approximate inference methods is novel as far as we know.  
It may also be possible to extend to adjustments beyond second order moments 
using the law of total cumulance \citep{brillinger69}, but we do not consider this further.  

\subsection{Use of a conditioning set}

A first observation (related to a similar remark in \cite{prangle+bps14} for their quantile based diagnostic for likelihood-free inference algorithms) 
is that while achieving approximate equality in (\ref{apmean}) and (\ref{apcov}) is necessary for an approximate inference algorithm to be good, 
it is not sufficient.  Consider, for example, the
approximate inference method which always returns the prior as the estimated posterior.  Then $E(\widetilde{E}(\thetavec|\Yvec))=E(E(\thetavec))=E(\thetavec)$, and
$\text{Cov}(\thetavec)=E(\widetilde{\text{Cov}}(\thetavec|\Yvec))+\text{Cov}(\widetilde{E}(\thetavec|\Yvec))=E(\text{Cov}(\thetavec))+\text{Cov}(E(\thetavec))=\text{Cov}(\thetavec)$ 
so (\ref{apmean}) and (\ref{apcov}) hold exactly if we always estimate the posterior regardless of $\yvec$ by the prior.  On the other hand, it is not hard to see
that diagnosing the quality of the approximation using (\ref{apmean}) and(\ref{apcov}) could be useful.  For example, suppose that our approximate inference
algorithm always produces exact posterior mean values $E(\thetavec|\Yvec)=\widetilde{E}(\thetavec|\Yvec)$ while underestimating variability $\text{Cov}(\thetavec|\Yvec)\geq \widetilde{\text{Cov}}(\thetavec|\Yvec))$
where $\Amat\geq \Bmat$ for covariance matrices $\Amat$ and $\Bmat$ means that $\Amat-\Bmat$ is non-negative definite.  Then (\ref{apmean}) would hold, but 
\begin{align*}
 \text{Cov}(\thetavec) & =  E(\text{Cov}(\thetavec|\Yvec))+\text{Cov}(E(\thetavec|\Yvec))  \geq E(\widetilde{\text{Cov}}(\thetavec|\Yvec))+\text{Cov}(\widetilde{E}(\thetavec|\Yvec)), 
\end{align*}
and the situation we have considered is not unrealistic for some variational inference algorithms where point estimation may be excellent but posterior variability is substantially 
underestimated.  

In the case of checking and adjustment of a posterior density $p(\thetavec|\yvec_{\text{obs}})$ for observed value $\yvec_{\text{obs}}$ of $\yvec$, it seems sensible to consider 
(again following \cite{prangle+bps14}) a conditioning on $\yvec\in F$ in the joint model for some set $F$ with $\yvec_{\text{obs}}\in F$ in order to make the adjustment
more relevant to the data observed.  
This will also avoid problems such as the one mentioned above where an approximate inference algorithm returning the prior would be considered satisfactory or in
no need of adjustment.  This is because after modifying equations  (\ref{apmean}) and (\ref{apcov}) by conditioning on $\yvec\in F$, these modified relations will not hold, in general, for an approximate inference algorithm estimating the posterior density by the prior.
The choice of $F$ will be discussed later in the examples.  More explicitly, consider the joint model 
\begin{align*}
  p^F(\thetavec,\yvec)\propto p(\thetavec)p(\yvec|\thetavec)\mathbf{1}(\yvec\in F),
\end{align*}
where $\mathbf{1}(\cdot)$ denotes the indicator function, 
and samples from this joint model can be obtained by drawing $(\thetavec',\yvec')\sim p(\thetavec)p(\yvec|\thetavec)$ until $\yvec '\in F$.  Write $p^F(\yvec)$ for the $\yvec$ marginal
of $p^F(\thetavec,\yvec)$.  In the model $p^F(\thetavec,\yvec)$ we can consider the following analogues of (\ref{priormean}) and (\ref{priorcov}):
\begin{align}
  E(\thetavec|\Yvec \in F) & = E(E(\thetavec|\Yvec)|\Yvec \in F), \label{pmeanA}
\end{align}
and
\begin{align}
  \text{Cov}(\thetavec|\Yvec \in F) & = E(\text{Cov}(\thetavec|\Yvec)|\Yvec\in F)+\text{Cov}(E(\thetavec|\Yvec)|\Yvec\in F),  \label{pcovA}
\end{align}
and then checking and adjustment can be based on assessment of departures from equality 
in (\ref{pmeanA}) and (\ref{pcovA}) when $E(\thetavec|\Yvec)$ and $\text{Cov}(\thetavec|\Yvec)$ are
replaced by $\widetilde{E}(\thetavec|\Yvec)$ and $\widetilde{\text{Cov}}(\thetavec|\Yvec)$.  If our approximation is accurate, then
\begin{align}
  E(\thetavec|\Yvec\in F) \approx & E(\widetilde{E}(\thetavec|\Yvec)|\Yvec\in F), \label{apmeanA}
\end{align}
and
\begin{align}
  \mbox{Cov}(\thetavec|\Yvec\in F) & \approx E(\widetilde{\text{Cov}}(\thetavec|\Yvec)|\Yvec\in F)+\text{Cov}(\widetilde{E}(\thetavec|\Yvec)|\Yvec\in F), \label{apcovA}
\end{align}
for samples from $p^F(\thetavec,\yvec)$.  In the next section we describe in detail how checking and adjustment are done when approximate posterior distributions are given
as particle approximations in the form of Monte Carlo samples.

\section{Checking and adjustment strategy}\label{sec:assessing}

For an approximate inference algorithm leading to approximate posterior distributions $\tilde{p}(\thetavec|\Yvec)$ for $\Yvec\sim p^F(\yvec)$
we have suggested checking and adjustment based on assessing departure from
equality in (\ref{apmeanA}) and (\ref{apcovA}).  For data $\yvec$ we assume that the approximation $\tilde{p}(\thetavec|\yvec)$ comes in the form of an approximate
posterior sample, $\thetavec^{(1)}(\yvec),\dots, \thetavec^{(S)}(\yvec)$, and if this is not the case, we assume that it is easy to generate from $\tilde{p}(\thetavec|\yvec)$ to obtain
such a sample.  Let $(\thetavec^{(i)},\yvec^{(i)})$, $i=1,\dots I$ denote independent draws from $p^F(\thetavec,\yvec)$.  

Let us write
\begin{align*}
  \muvec^L & = \frac{1}{I} \sum_{i=1}^I \thetavec^{(i)} \hspace{0.25in}\text{and}\hspace{0.25in}
  \muvec^R = \frac{1}{I} \sum_{i=1}^I \muvec^R(\yvec^{(i)}),
\end{align*}
with 
\begin{align*}
  \muvec^R(\yvec^{(i)}) & = \frac{1}{S}\sum_{s=1}^S \thetavec^{(s)}(\yvec^{(i)}).
\end{align*}
Here $\muvec^L$ and $\muvec^R$ are sample based estimates of the left- and right-hand sides of (\ref{apmeanA}).  
Also write
\begin{align*}
 \Sigmamat^L & = \frac{1}{I-1} \sum_{i=1}^I (\thetavec^{(i)}-\muvec^L)(\thetavec^{(i)}-\muvec^L)^T,
\end{align*}
and
\begin{align*}
  \Sigmamat^R & = \Sigmamat^{R1}+\Sigmamat^{R2},
\end{align*}
where
\begin{align*}
 \Sigmamat^{R1} & = \frac{1}{I} \sum_{i=1}^I \Sigmamat^{R1}(\yvec^{(i)}),
\end{align*}
\begin{align*}
  \Sigmamat^{R1}(\yvec^{(i)}) & = \frac{1}{S-1}\sum_{s=1}^S (\thetavec^{(s)}(\yvec^{(i)})-\muvec^R(\yvec^{(i)}))(\thetavec^{(s)}(\yvec^{(i)})-\muvec^R(\yvec^{(i)}))^T,
\end{align*}
and
\begin{align*}
 \Sigmamat^{R2} & = \frac{1}{I-1} \sum_{i=1}^I (\muvec^R(\yvec^{(i)})-\muvec^R)(\muvec^R(\yvec^{(i)})-\muvec^R)^T.
\end{align*}
Here $\Sigmamat^L$ and $\Sigmamat^R$ are sample estimates of the left- and right-hand sides of (\ref{apcovA}), with 
$\Sigmamat^{R1}$ and $\Sigmamat^{R2}$ sample estimates of the first and second terms on the right-hand side of (\ref{apcovA}).  
In the case where $\widetilde{E}(\thetavec|\yvec^{(i)})$ and $\widetilde{\text{Cov}}(\thetavec|\yvec^{(i)})$ are known exactly, these can be substituted
for $\muvec^R(\yvec^{(i)})$ and $\Sigmamat^R(\yvec^{(i)})$ above.  This can happen when the approximation is not in the form of a Monte Carlo sample 
but takes some analytical form such as for a variational approximation.  With the above ideas and notation established, we consider
next inference algorithm assessment, and then inference adjustments.  

\subsection{Assessment of the inference algorithm}

The equalities $\muvec^L=\muvec^R$ and $\Sigmamat^L=\Sigmamat^R$ will not hold exactly.  We suggest assessing whether components of $\muvec^L$ and $\muvec^R$ and of $\Sigmamat^L$ and $\Sigmamat^R$ differ by a large amount compared to the variability in their sample based estimation.  The practical importance of the size of any difference for inferences and decisions should also be considered, but this question tends to be application specific.  To assess variability in sample based estimation of $\muvec^L$, $\muvec^R$, $\Sigmamat^L$ and $\Sigmamat^R$ we suggest using the bootstrap. 
Consider resampling of the triples $(\thetavec_i,\muvec^R(\yvec^{(i)}),\Sigmamat^{R1}(\yvec^{(i)}))$, $i=1, \dots, I$, with replacement, $B$ times.  For each resample we compute $\muvec^L$, $\muvec^R$, $\Sigmamat^L$ and $\Sigmamat^R$ to obtain values $\muvec^{Lb}$, $\muvec^{Rb}$, $\Sigmamat^{Lb}$, $\Sigmamat^{Rb}$, $b=1,\dots, B$.  We can plot components of $\muvec^{Lb}$ and $\muvec^{Rb}$ against each other, as well as standard deviations and correlations derived from $\Sigmamat^{Lb}$ and $\Sigmamat^{Rb}$.   Estimates $\muvec^L$ and $\Sigmamat^L$ are direct estimates of the mean and covariance of $\thetavec$ given $\Yvec\in F$ based on prior samples.  Estimates  $\muvec^R$ and $\Sigmamat^R$ are indirect estimates based on posterior approximations.   
For assessment of an approximate inference algorithm, checks of location, scale and correlation are often directly meaningful to users.  In contrast, marginal quantile based assessments require the user to back out information about the nature
of any miscalibration in location and scale from histograms of $p$-values, which of course can be done but requires more sophistication. 
For our suggested plots of functions of $\muvec^{LB}, \muvec^{Rb}$ and $\Sigmamat^{Lb},\Sigmamat^{Rb}$ against each other, points lying above (below)
the diagonal line indicate an average overestimation (underestimation) based on the posterior approximation.
The use and meaning of these plots is discussed further in the examples of Section 4.

\subsection{Adjustment of the inference algorithm}

To adjust approximate inferences so that the equalities (\ref{apmeanA}) and (\ref{apcovA}) hold,   we can transform the
particles
$\thetavec^{(s)}(\yvec^{(i)})$.  We assume that the components of $\thetavec$ are unrestricted, which can be achieved by a preliminary
transformation if necessary.
Write $\Cmat$ for the lower triangular Cholesky factor of $\Sigmamat^{R1}$.  Also, 
assuming that $\Sigmamat^L-\Sigmamat^{R2}$ is positive definite, write $\Tmat$ for it's lower triangular Cholesky factor.  
Although $\Sigmamat^L-\Sigmamat^{R2}$ is not guaranteed to be positive definite in general, it usually is, since the prior variation (captured by $\Sigmamat^L$) tends to be
larger than the variability in the posterior means (estimated by $\Sigmamat^{R2}$).  
We describe later how to handle the situation where $\Sigmamat^L-\Sigmamat^{R2}$ is not positive definite, which
can which can happen when there is a very poor approximation or because of the sampling variability in the estimation
of $\Sigmamat^L$ and $\Sigmamat^{R2}$.  

Now suppose we transform the samples $\thetavec^{(s)}(\yvec^{(i)})$, $i=1,\dots, I$, $s=1,\dots, S$ to
\begin{align}
  \tilde{\thetavec}^{(s)}(\yvec^{(i)})=\muvec^L+(\muvec^R(\yvec^{(i)})-\muvec^R)+\Tmat \Cmat^{-1}(\thetavec^{(s)}(\yvec^{(i)})-\muvec^R(\yvec^{(i)})). \label{transformedsamples}
\end{align}
The first two terms on the right-hand side of (\ref{transformedsamples}) perform a mean adjustment for the $i$th replicate samples sufficient to ensure
that (\ref{apmeanA}) holds for the adjusted samples.  The final term on the right-hand side of (\ref{transformedsamples}) is a scaling of the mean-centred particles
for the $i$th replicate samples, and ensures that an empirical estimate of $E(\text{Cov}(\theta|Y))$ based on the transformed particles leads to equality in 
(\ref{apcovA}).    
More precisely, denote the quantities $\muvec^R(\yvec^{(i)}), \muvec^R,\Sigmamat^{R1}(\yvec^{(i)})$ etc.~evaluated for the adjusted samples
by $\tilde{\muvec}^R(\yvec^{(i)}),\tilde{\muvec}^R,\tilde{\Sigmamat}^{R1}(\yvec^{(i)})$, etc.  We have
\begin{align*}
\tilde{\muvec}^R & = \frac{1}{I}\sum_{i=1}^I \tilde{\muvec}^R(\yvec^{(i)}),
\end{align*}
with 
\begin{align*}
\tilde{\muvec}^R(\yvec^{(i)}) & = \frac{1}{S} \sum_{s=1}^S \tilde{\thetavec}^{(s)}(\yvec^{(i)})=\muvec^L+(\muvec^R(\yvec^{(i)})-\muvec^R), 
\end{align*}
so that $\tilde{\muvec}^R=\muvec^L$ and (\ref{apmeanA}) holds exactly for the adjusted samples.  
Next, note that 
\begin{align*}
\tilde{\muvec}^R(\yvec^{(i)})-\tilde{\muvec}^R & =\muvec^R(\yvec^{(i)})-\muvec^R,
\end{align*}
and hence $\tilde{\Sigmamat}^{R2}=\Sigmamat^{R2}$.  
Also,  
$\tilde{\thetavec}^{(s)}(\yvec^{(i)})-\tilde{\muvec}^R(\yvec^{(i)})=\Tmat \Cmat^{-1}(\thetavec^{(s)}(\yvec^{(i)})-\muvec^R(\yvec^{(i)})),$ and hence
\begin{align*}
 \tilde{\Sigmamat}^{R1}&=\frac{1}{I}\sum_{i=1}^I \Tmat \Cmat^{-1} \Sigmamat^{R1}(\yvec^{(i)})\Cmat^{-\top}\Tmat^\top=\Tmat \Cmat^{-1} \Sigmamat^{R1} \Cmat^{-\top} \Tmat^\top=\Tmat \Tmat^\top=\Sigmamat^L-\Sigmamat^{R2},
\end{align*}
so that $\tilde{\Sigmamat}^{R1}=\Sigmamat^L-\Sigmamat^{R2}$ and 
\begin{align*}
  \tilde{\Sigmamat}^R & = \tilde{\Sigmamat}^{R1}+\tilde{\Sigmamat}^{R2}= \Sigmamat^L-\Sigmamat^{R2}+\Sigmamat^{R2}=\Sigmamat^L,
\end{align*}
so (\ref{apcovA}) also holds exactly for the adjusted samples.  
Given samples $\thetavec^{(s)}(\yvec_{\text{obs}})$, $s=1,\dots, S$ for the observed data, we will transform them similarly to (\ref{transformedsamples}) to 
get an adjusted Monte Carlo sample from the posterior given $\yvec_{\text{obs}}$ to be used for inferential purposes:
\begin{align}
 \tilde{\thetavec}^{(s)}(\yvec_{\text{obs}}) & = \muvec^L+(\muvec^R(\yvec_{\text{obs}})-\muvec^R)+\Tmat \Cmat^{-1}(\thetavec^{(s)}(\yvec_{\text{obs}})-\muvec^R(\yvec_{\text{obs}})). \label{adjpostsamples}
\end{align}

As mentioned earlier it may happen that $\Sigmamat^L-\Sigmamat^{R2}$ is
not positive definite when the approximate inference algorithm gives particularly poor estimation
of the posterior mean values, or due to sampling variability in the estimates $\Sigmamat^L$ and $\Sigmamat^{R2}$.  
To handle this we can do a preliminary preprocessing in which $\thetavec^{(s)}(\yvec^{(i)})$ is changed to 
\begin{eqnarray*}
 & \muvec^R+\sqrt{\rho}(\muvec^R(\yvec^{(i)})-\muvec^R)+\thetavec^{(s)}(\yvec^{(i)})-\muvec^R(\yvec^{(i)}),
\end{eqnarray*}
in which $\rho$ is a shrinkage parameter, $0<\rho<1$.  
It is easy to see that this results in shrinkage of the $\muvec^R(\yvec^{(i)})$ for the new particles towards $\muvec^R$, with $\Sigmamat^{R2}$ changing to $\rho \Sigmamat^{R2}$.  The preprocessing leaves 
$\muvec^R$, $\Sigmamat^{R1}(\yvec^{(i)})$ and $\Sigmamat^{R1}$ unchanged.  The parameter $\rho$ is chosen so that $\Sigmamat^L-\rho \Sigmamat^{R2}$ is
positive definite.  We suggest choosing it so that the smallest eigenvalue of $\Sigmamat^L-\rho \Sigmamat^{R2}$ is equal to the
smallest eigenvalue of $\Sigmamat^{R1}$.  Once this preprocessing is done if necessary, our adjustment can proceed as above.  

The adjustment suggested (and the similar quantile based adjustment and checking methods in the literature) are computationally intensive, since they 
require repeated approximations to the posterior for different datasets.  However, these approximations are sometimes easy to obtain.  For example, in the 
case of approximate Bayesian computation methods using samples from the prior, the same prior samples can be reused for the approximation for different data.  
In the case where fast enough posterior approximations are available \cite{rodrigues+ps17} point out that recalibration adjustments can still be attractive.  
The computations involved in recalibration are trivially parallelizable, consisting of independent computations 
for different datasets.  

\section{Applications}\label{sec:applications}

We consider three examples.  The first concerns a likelihood-free inference application discussed in \cite{rodrigues+ps17}, and we compare our own adjustment method with the quantile-based method considered in their paper.  
Our second example involves inference in a deep neural network generalized linear mixed model \citep{tran+nnk18}.  
In this example our method corrects for empirical Bayes and variational computational approximations
in approximate Bayesian inference.  We also explain the features of this example which make 
the adjustment of \cite{rodrigues+ps17} performs poorly compared to our new method.
The third example considers predictive inference in a Gaussian process model, when a deep neural network regression
is used as a surrogate for the Gaussian process.  
The inference adjustment is intended to make the deep neural network regression
uncertainty quantification closer to that provided by the Gaussian process.  This can be of interest because 
the Gaussian process computation can be intractable for large datasets.

\subsection{Likelihood-free inference using an auxiliary model}\label{subsec:likefree}

Our first example was considered in \cite{rodrigues+ps17}, 
where they considered a simple model that is 
useful in some finance and insurance applications.  
Consider $n$ independently and identically distributed observations 
$\yvec=(y_1,\dots, y_n)$ of a random variable $Y$ that is defined as
$Y=\sum_{k=1}^\kappa W_k$ where the $W_k$ are independent log-normal, $W_k\sim \text{LogNormal}(\mu,\sigma^2)$.  The density of $Y$ is difficult to compute,
and \cite{rodrigues+ps17} consider the so-called Fenton-Wilkinson approximation \citep{fenton60,asmussen+r08} which uses a log-normal density matching the mean and variance
of $Y$ to approximate the intractable density.  $Y$ is approximated by $U\sim \text{LogNormal}(m,s^2)$, where
\begin{align*}
  m & = \mu+\log \kappa+\frac{1}{2}(\sigma^2-s^2) \\
 s^2 & = \log \left\{(\exp(\sigma^2)-1)/\kappa+1\right\}.
\end{align*}
As before, the observed $\yvec$ is denoted $\yvec_{\text{obs}}$.  As a fast auxiliary model approximation to the posterior
density for the parameter $\thetavec=(\mu,\eta=\log \sigma^2)$,
\cite{rodrigues+ps17} consider a Laplace approximation when the likelihood is approximated by the Fenton-Wilkinson method.  For data $\yvec$, write
$\tilde{p}(\yvec|\thetavec)$ for the approximate likelihood, $\hat{\thetavec}(\yvec)=(\hat{\mu}(\yvec),\hat{\eta}(\yvec))$ for the mode of $\log h(\thetavec)$ where
$h(\thetavec)=p(\thetavec)\tilde{p}(\yvec|\thetavec)$, and $H(\hat{\thetavec}(\yvec))$ for the Hessian of $-\log h(\thetavec)$ evaluated at $\hat{\thetavec}(\yvec)$.  
The posterior approximation based on the auxiliary Fenton-Wilkinson model and Laplace approximation 
is $N(\hat{\thetavec}(\yvec),H(\hat{\thetavec}(\yvec))^{-1})$.  
Computation of this approximation is very fast, whereas an ABC approximation using $\hat{\thetavec}(\yvec)$ as the summary statistic is computationally demanding.  After obtaining a posterior approximation for $\thetavec$, we transform back to an approximation
for $(\mu,\sigma)$.
Although this example is concerned with likelihood-free inference, the use of an auxiliary model is not the traditional ABC approach commonly used in such settings.  
For ABC, there are a number of other adjustment methods in the literature.  These include methods based on regression 
\citep{beaumont+zb02,blum10,blum+t10,blum+f10} and on making adjustments to low-dimensional marginals \citep{nott+fms14,li+nfs15}.

\cite{rodrigues+ps17} consider simulating $n=10$ observations with $\kappa=10$ and $(\mu,\sigma)=(0,1)$.  The prior densities
on $\mu$ and $\sigma$ are independent $N(0,1)$ and $\text{Gamma(1,1)}$ respectively.
The recalibration procedure in Algorithm 2 of \cite{rodrigues+ps17} incorporates
kernel weights.  With a uniform kernel, this corresponds to choosing a conditioning set $F$ in their approach, similar to that considered in Section 2, 
based on the support of the kernel centred on 
$\hat{\thetavec}(\yvec_{\text{obs}})$.  In this example they use kernel bandwidth $h=\infty$, so that there is no reweighting or conditioning.  
As starting point for their adjustment, they
generate a training sample of $10,000$ parameter and data pairs from the prior.  For each of these they compute their auxiliary model posterior approximation.
Our own multivariate adjustment approach will be implemented based on these same posterior approximations.
For our new method, we use a conditioning set $F$ of the form
\begin{align*}
  F & = \{y: w_1^{-2}(\hat{\mu}(\yvec)-\hat{\mu}(\yvec_{\text{obs}}))^2+w_2^{-2}(\hat{\eta}(\yvec)-\hat{\eta}(\yvec_{\text{obs}}))^2<c\},
\end{align*}
where 
$w_1$ and $w_2$ are the prior predictive mean absolute deviations of $\hat{\mu}(\yvec)$ and $\hat{\sigma}(\yvec)$ respectively estimated
from the training sample, and $c$ is chosen so that $1,000$ of
the $10,000$ prior training samples are covered by $F$.  

\begin{figure}
\centering
\begin{tabular}{c}
    \includegraphics[scale=0.75]{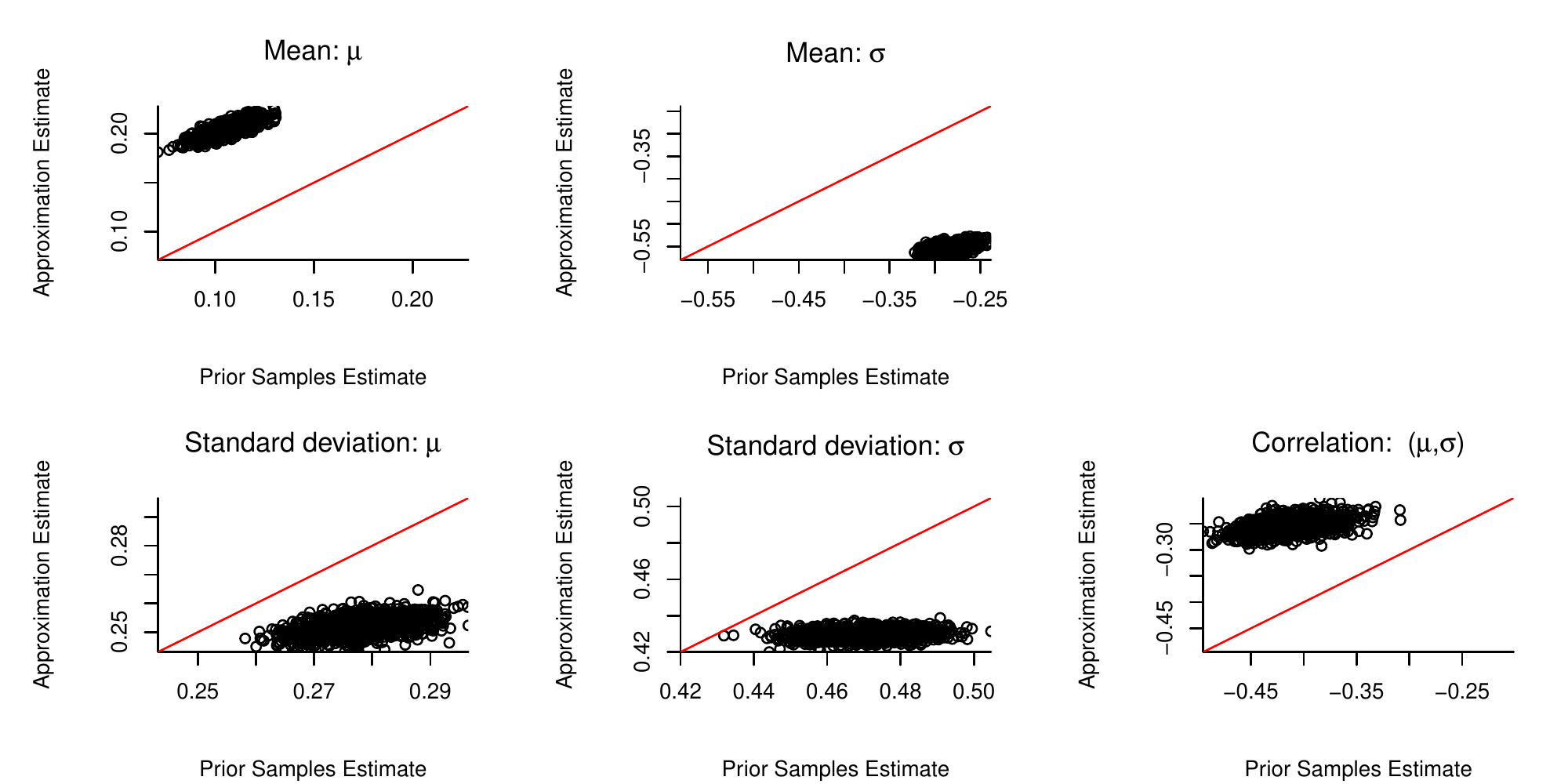} 
\end{tabular}
  \caption{\label{fig1} Plots of bootstrap replicates of components of $\muvec^L$ and $\muvec^R$ and of standard deviations and correlations derived from $\Sigmamat^L$ and $\Sigmamat^R$ against each other, with $B=1000$.  
Points above the red diagonal line indicate an overall overestimation by the posterior approximation, points below the red diagonal line indicate an overall underestimation. }
\end{figure}

First, we consider a check on the quality of the approximate inferences provided by the auxiliary model.  Figure \ref{fig1} shows plots of the kind suggested
in Section 3.1 of the components of bootstrap replicates of $\muvec^{Lb}$ and $\muvec^{Rb}$ against each other, and of posterior standard deviations and correlations
derived from bootstrap replicates of $\Sigmamat^{Lb}$ and $\Sigmamat^{Rb}$.  The values for $\muvec^L$ and $\Sigmamat^L$ are direct estimates of moments for $\thetavec$ based
on prior samples - the values for $\muvec^R$ and $\Sigmamat^R$ are indirect estimates of moments based on the posterior approximation.  
The plots suggest some overall overestimation of posterior means for $\mu$, underestimation of posterior means for $\sigma$, underestimation of standard deviations
for $\mu$ and $\sigma$, and overestimation of correlation between $\mu$ and $\sigma$.

\begin{figure}
\centering
\begin{tabular}{c}
    \includegraphics[scale=0.6]{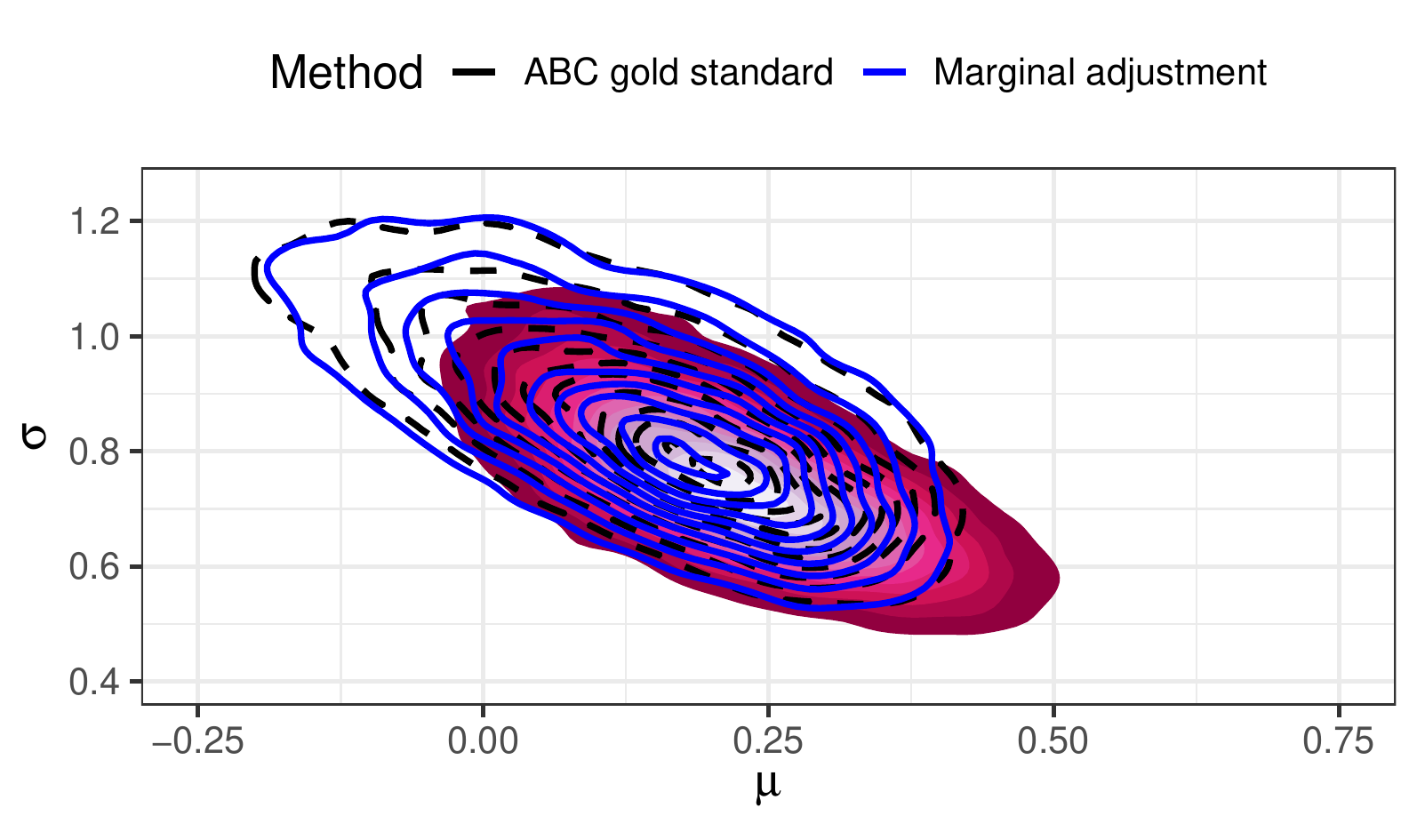} \\
    \includegraphics[scale=0.6]{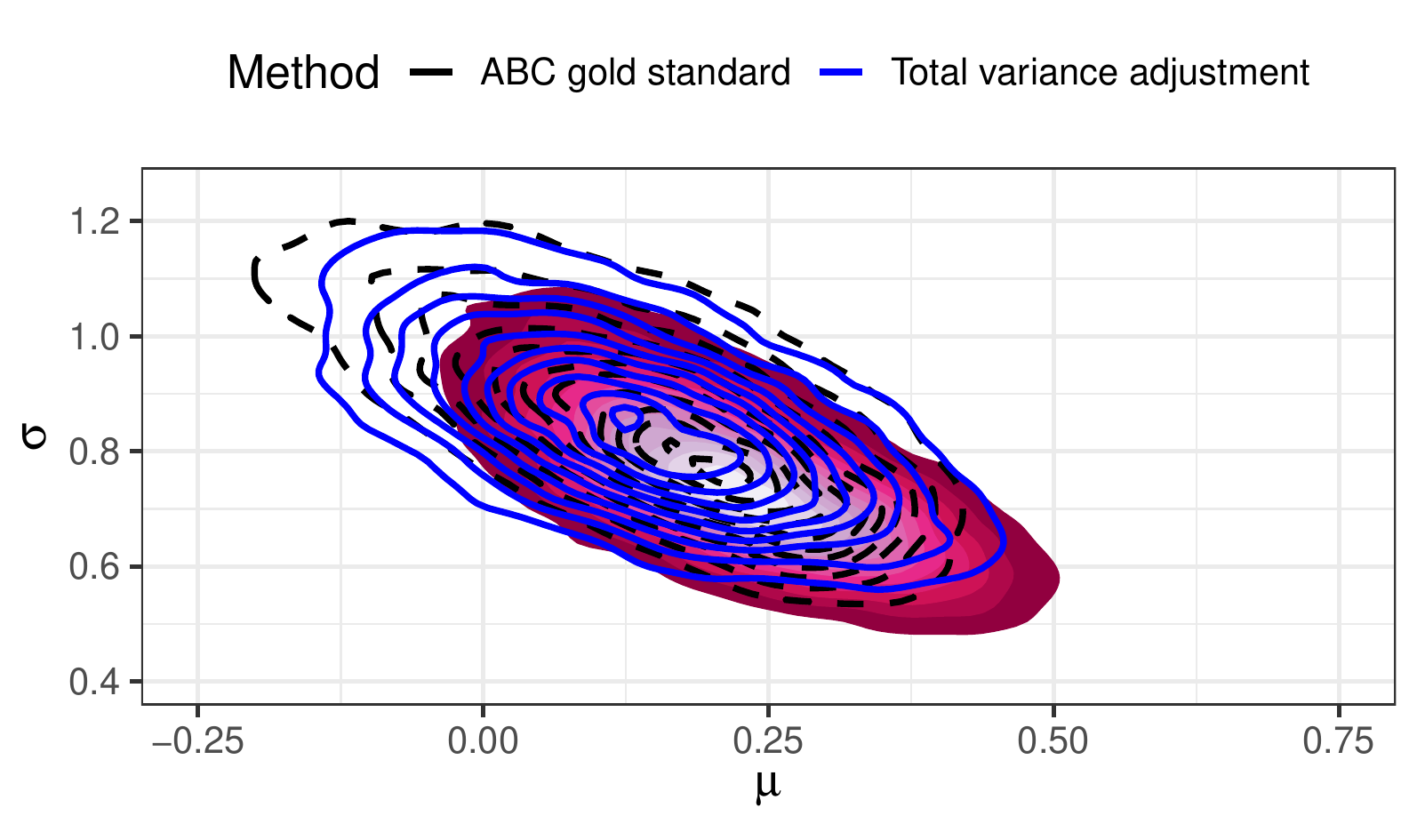} 
\end{tabular}
  \caption{\label{fig2} Contour plots of uncalibrated approximation, ``gold standard" ABC approximation and adjusted approximations based on marginal calibration (top) and the new total variance calibration (bottom).  All posterior approximations shown are two-dimensional kernel density estimates obtained from particle approximations to the posterior density for the different methods. }
\end{figure}

The method of \cite{rodrigues+ps17} works well in this example.  The top panel of Figure \ref{fig2} is similar to Figure 1 (b) in \cite{rodrigues+ps17} and 
shows the estimated joint posterior based on the adjusted samples for 
their method, compared to a ``gold standard" ABC method with small tolerance (see \cite{rodrigues+ps17} for further details).  The filled contours show the 
unadjusted auxiliary approximation.  Shown in the bottom panel of Figure \ref{fig2}
is the adjusted posterior density estimate for our new approach, again comapred with the ``gold standard" ABC analysis and the unadjusted approximation.  
Since the new method performs only mean and scale adjustments, the estimated joint posterior for $\thetavec$ will be normal after adjustment when
the auxiliary model posterior approximation is normal.  Figure \ref{fig2} shows that both methods manage to correct for 
the miscalibration of mean and scale in the initial auxiliary model approximation.  The method of \cite{rodrigues+ps17} works best here, 
but we explain in the next example why it sometimes performs badly.

\subsection{Variational and empirical Bayes approximations within a deep longitudinal mixed model}\label{subsec:deeplearning}

\citet{tran+nnk18} describe generalized linear mixed models (GLMMs) for longitudinal data using deep neural network (DNN)
basis functions.  We consider inference for random effects variances in such models.  Write $y_{j,t}$
for a binary response for an individual $j$ at time $t$, $j=1,\dots, J$, $t=1,\dots, T_j$, and write
$\xvec_{j,t}$ for a corresponding vector of covariates.  A logistic deep GLMM assumes
\begin{align*}
  y_{j,t}\sim \text{Bernoulli}(p_{j,t}),\qquad \log\left(\frac{p_{j,t}}{1-p_{j,t}}\right)=\mathfrak{N}(\xvec_{j,t},\wvec,\betavec+\alphavec_j),
\end{align*}
where $\mathfrak{N}(\xvec_{j,t},\wvec,\betavec+\alphavec_j)$ is the scalar output of a feed-forward neural network with inner weights $\wvec$ and
output weights $\betavec+\alphavec_j$, for input $\xvec_{j,t}$.  If $L$ is the number of nodes in the last hidden layer of the neural
network, the dimension of $\betavec$ and each $\alphavec_j$ is $L+1$, after inclusion of an intercept term.  
The vector $\betavec$ contains fixed effects parameters and $\alphavec_j$ are random effects, 
modelled as $\alphavec_j\sim N(\zerovec,\Gammamat)$ where $\Gammamat$ is diagonal with diagonal entries $\Gamma_i$, $i=0,\dots, L$, and
$\Gamma_0$ is the variance of the random intercepts.
In our later example $L=10$, and only a single hidden layer will be used in the neural network.  
The model parameters are $\thetavec=(\wvec,\betavec,\Gamma_0,\dots, \Gamma_{L})^\top$. 
Semiparametric versions of the above model which are more interpretable and in which some of the covariates are not transformed can also 
be considered.  See \citet{tran+nnk18} for further discussion of the model.

For the neural network weights we use a prior which is normal, 
$\wvec\sim N(\zerovec,\frac{1}{\gamma}\Imat)$, $\beta\sim N(\zerovec,\frac{1}{\gamma}\Imat)$, where
in our later data analysis $\gamma=130$.  The value $\gamma=130$ was chosen based
on prior predictive simulations.  The parameters $\Gamma_i$ $i=0,\dots, L$ 
are assumed to follow a prior in which they are independent with $\Gamma_{i}\sim \text{Gamma}(a_0,b_0)$, 
but constrained to the region $\Gamma_{1}>\Gamma_2>\dots >\Gamma_L$.  This means that the basis functions
are ordered according to the size of the corresponding random effect variance.  
We set $a_0=1$ and $b_0=10$ in this example, where the shape-scale parametrization of the gamma distribution is used here.  
We focus on inference for $\thetavec=(\log \Gamma_1,\dots, \log \Gamma_{L})^\top$.  
It is challenging to make inferences about a large set of random effects variance parameters here unless
the number of subjects is large.  
In generalized linear mixed models, maximum likelihood estimates or posterior mode estimates
with weak priors often lead to degenerate estimates \citep{chung+grld15}.
Here we use strong shrinkage priors and a random effect for each basis function to define a predictive
model which flexibly describes within subject dependence.  

We use our adjustment method to approximate full Bayes inference using the normal prior on the
weights above, based on the approximate normal posterior produced by the algorithm of \citet{tran+nnk18}.  
\citet{tran+nnk18} approximate Bayesian inference under a different prior specification for the weights
and using variational approximation methods and empirical Bayes methods for hyperparameter estimation.  
See Section 6.1.4 of their paper.
\citet{tran+nnk18} do not impose the identification constraint $\Gamma_{1}>\dots, >\Gamma_L$ for inference, and in 
their normal posterior approximation it is assumed that the 
covariance matrix has the low-rank plus diagonal form $\Bvec\Bvec^\top+\Dmat^2$, where $\Dmat$ is diagonal and it will be assumed
here that $\Bvec$ is a column vector.  The approximate posterior samples needed for our adjustment method
are obtained by simulating from the normal variational posterior, and then reordering components so that
$\Gamma_1>\dots >\Gamma_L$.  We take the effect of the relabelling as another aspect of the approximation that
our method aims to correct.  

Similar to the previous example, we use a conditioning set $F$.  If $\yvec$ is a simulated dataset from
the prior with $\yvec=(\yvec_1^\top,\dots, \yvec_n^\top)^\top$, $\yvec_j=(y_{j,1},\dots, y_{j,T_j})^\top$, then we use
\begin{align*}
  F & = \left\{y:\text{dist}(\yvec,\yvec_{\text{obs}})<\epsilon\right\},
\end{align*}
where $\yvec_{\text{obs}}$ is the observed data, $\yvec_{\text{obs}}=(\yvec_{\text{obs},1}^\top,\dots, \yvec_{\text{obs},n}^\top)^\top$, 
$\yvec_{\text{obs},j}=(y_{\text{obs},j1},\dots, y_{\text{obs},jT_j})^\top$, $0<\epsilon<1$ is a tolerance parameter, and
\begin{align*}
  \text{dist}(\yvec,\yvec_{\text{obs}}) & = \frac{1}{n}\sum_{j=1}^n \text{dist}_{\text{Jaccard}}(\yvec_j,\yvec_{\text{obs},j}),
\end{align*}
with $\text{dist}_{\text{Jaccard}}(\uvec,\vvec)$ the Jaccard distance \citep{jaccard01} between two binary vectors with the same length.
If $a$ is the number of components where $\uvec$ and $\vvec$ are both $1$ and $b$ is the number of 
components where $\uvec$ and $\vvec$ take different values, then
$\text{dist}_{\text{Jaccard}}(\uvec,\vvec)=b/(a+b)$, where here by convention $0/0$ is defined to be $0$.
We use our adjustment method with $I=200$ prior replicates, and choose $\epsilon$ in the conditioning set so that $100$
of these prior replicates are kept for the adjustment.

We consider an  analysis of a German health care dataset \citep{geil+mrz97} consisting of yearly records of hospitalized status
of 1887 German workers.  The response $y_{j,t}=1$ if worker $j$ was hospitalized in year $t$, and $y_{j,t}=0$ otherwise.  
The covariates include gender, age, income, education, marriage status, job type and insurance type.  
There are 12 predictors in total after using dummy variables to represent the categorical variables.  
We use only the first $500$ panels in the dataset, since fitting the deep GLMM model is computationally expensive.  

\subsubsection{Assessment of approximate inference}

First, consider our bootstrap assessment plots for $\Deltavec=(\Delta_1,\Delta_2,\Delta_3,\Delta_4)^\top$, where
$\Delta_j=\log \Gamma_j$.
We focus on the four largest random effects variance parameters since they are the most important, and summarizing 
the results graphically is easier with a smaller number of parameters.
Figure 3 shows the plots for the mean (top row) and standard deviation (bottom row).  We see that 
the approximate inference method underestimates the largest $\log \Gamma_j$ values, and overestimates their variability.
\begin{figure}
\centering
\begin{tabular}{c}
\includegraphics[scale=0.5]{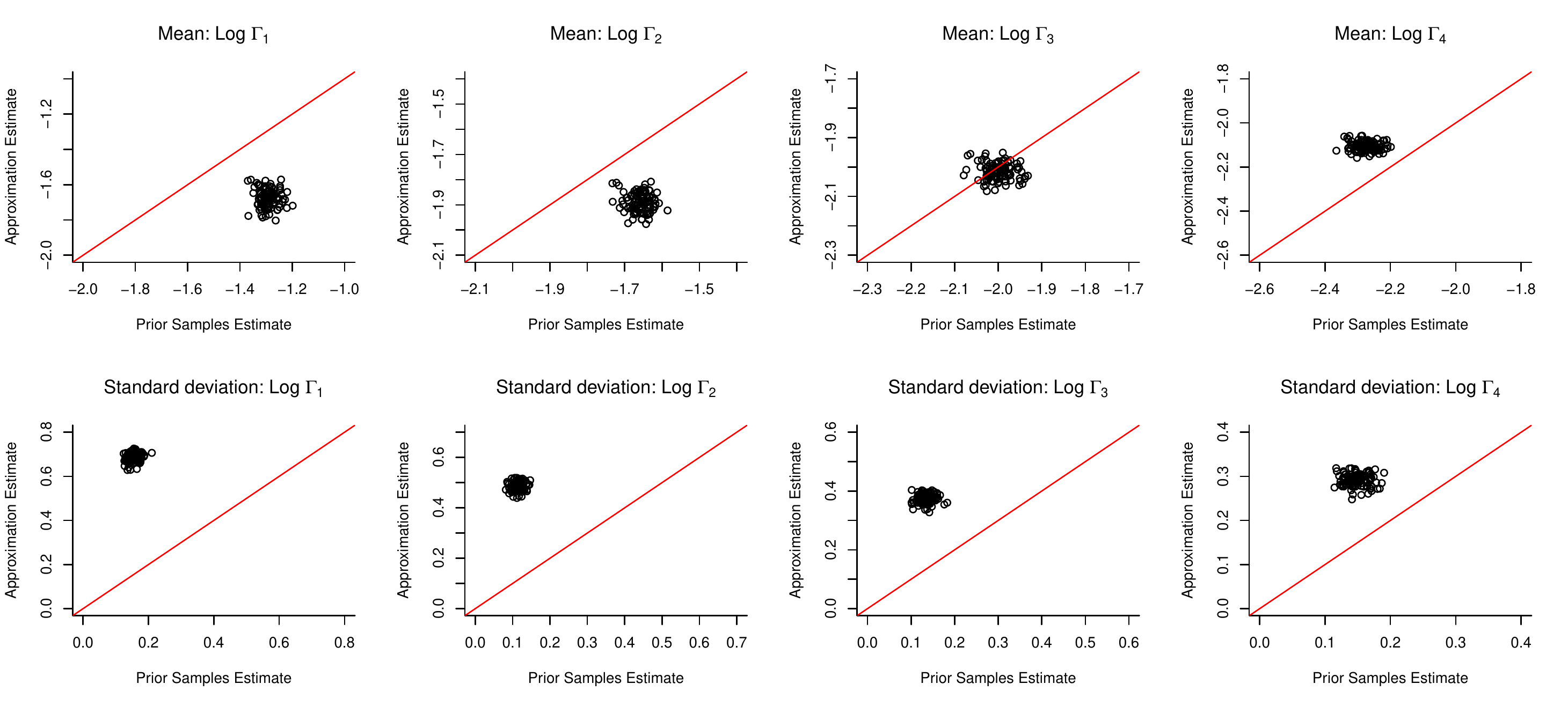} 
\end{tabular}
  \caption{\label{fig3} Plots of bootstrap replicates of components of $\muvec^L$ and $\muvec^R$ (top row) and of standard deviations 
 derived from $\Sigmamat^L$ and $\Sigmamat^R$ (bottom row) against each other for $(\log \Gamma_1,\log \Gamma_2,\log \Gamma_3,\log \Gamma_4)$.  Points above the diagonal line indicate an overall overestimation by the posterior approximation, points below the red diagonal line indicate an overall underestimation. }
\end{figure}

\subsubsection{Approximate inference adjustments}

When the value of $E(\Delta_j|y^{(i)})$ is estimated by $\hat{\Delta}_j^{(i)}$, the squared estimation error
is $|\Delta_j^{(i)}-\hat{\Delta}_j^{(i)}|^2$ where $\Delta_j^{(i)}$ is the value of $\Delta_j$ used in simulating 
the data for replicate $i$.  
Figure 4 shows boxplots of these squared estimation errors for estimated posterior means 
obtained from unadjusted and adjusted
posterior samples.   The total variance adjustment improves substantially on the unadjusted method.  
We also tried the adjustment method of \cite{rodrigues+ps17}, which does not work well in this example (results not shown).  
The reason is that the unadjusted posterior approximation is very poor here, and the method of \citet{rodrigues+ps17}
obtains adjusted sample components by evaluating marginal posterior quantile functions conditional on the observed data at certain points.  
If the quantile functions are estimated based on the empirical distribution of a particle posterior estimate, then the adjusted
samples cannot extend beyond the observed range of the unadjusted particles.  Hence, if a good approximation requires putting
posterior mass beyond the range of the unadjusted particles, the method of \citet{rodrigues+ps17} can perform badly.  
The total variance method can still give useful results here because of the way that adjustments are based on mean and 
scale adjustments rather than quantiles.
\begin{figure}
\centering
\begin{tabular}{c}
\includegraphics[scale=0.5]{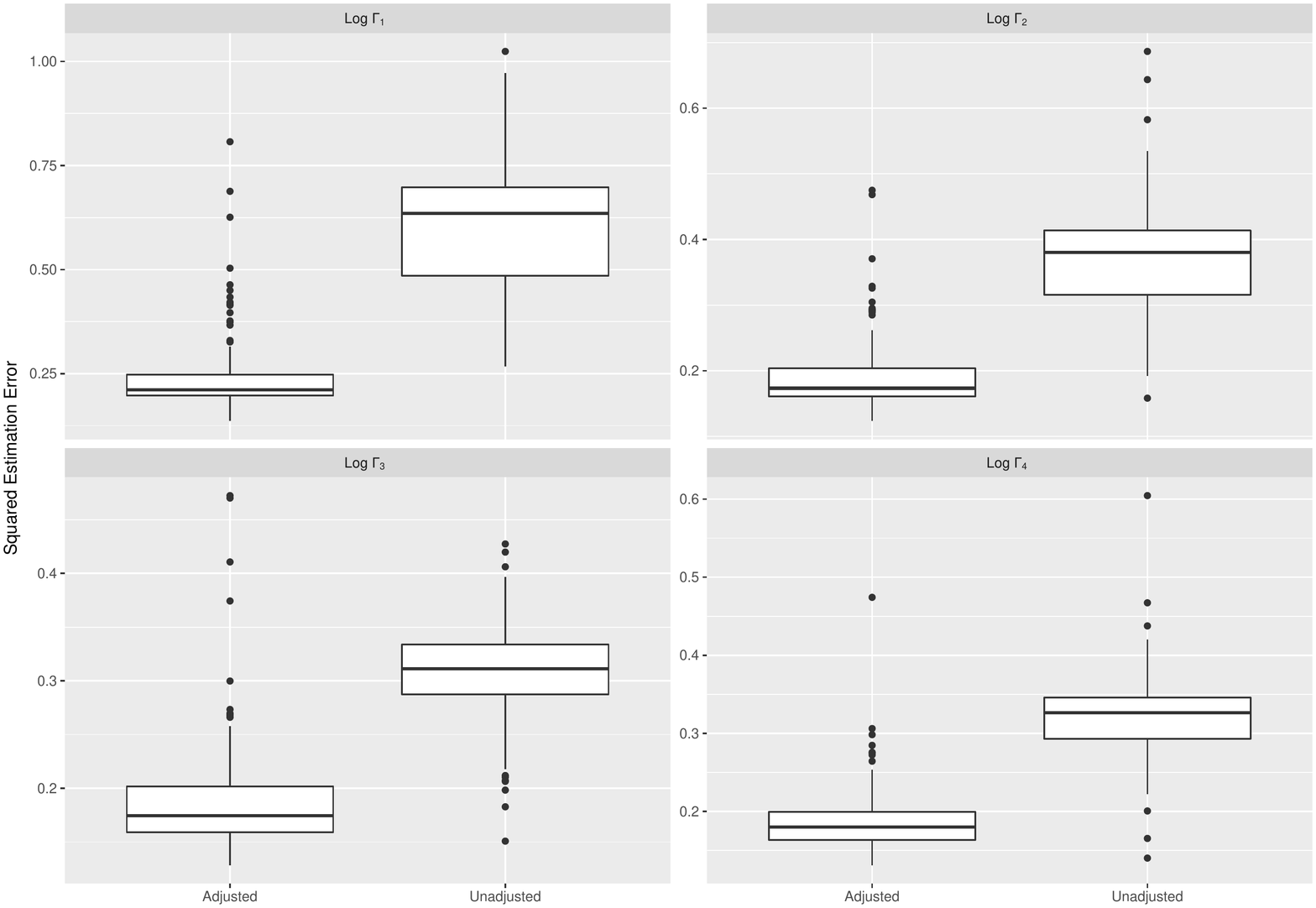} 
\end{tabular}
  \caption{\label{fig4} Boxplots of squared estimation errors for $\Delta_j=\log \Gamma_j$, $j=1,\dots, 4$, with and without adjustment, 
  treating each simulation replicate in turn as the observed data}
\end{figure}

Figure 5 shows, for the observed data, kernel estimates of the univariate and bivariate marginal posterior densities
obtained with and without the total variance adjustment.   Here we
do not know the true parameter values.  However, given
the improvements demonstrated in Figure 4 for the simulated data, inference based on the adjusted 
approach seems preferable, and furthermore this adjustment leads to substantially different estimated
posterior densities.
\begin{figure}
\centering
\begin{tabular}{c}
\includegraphics[scale=0.5]{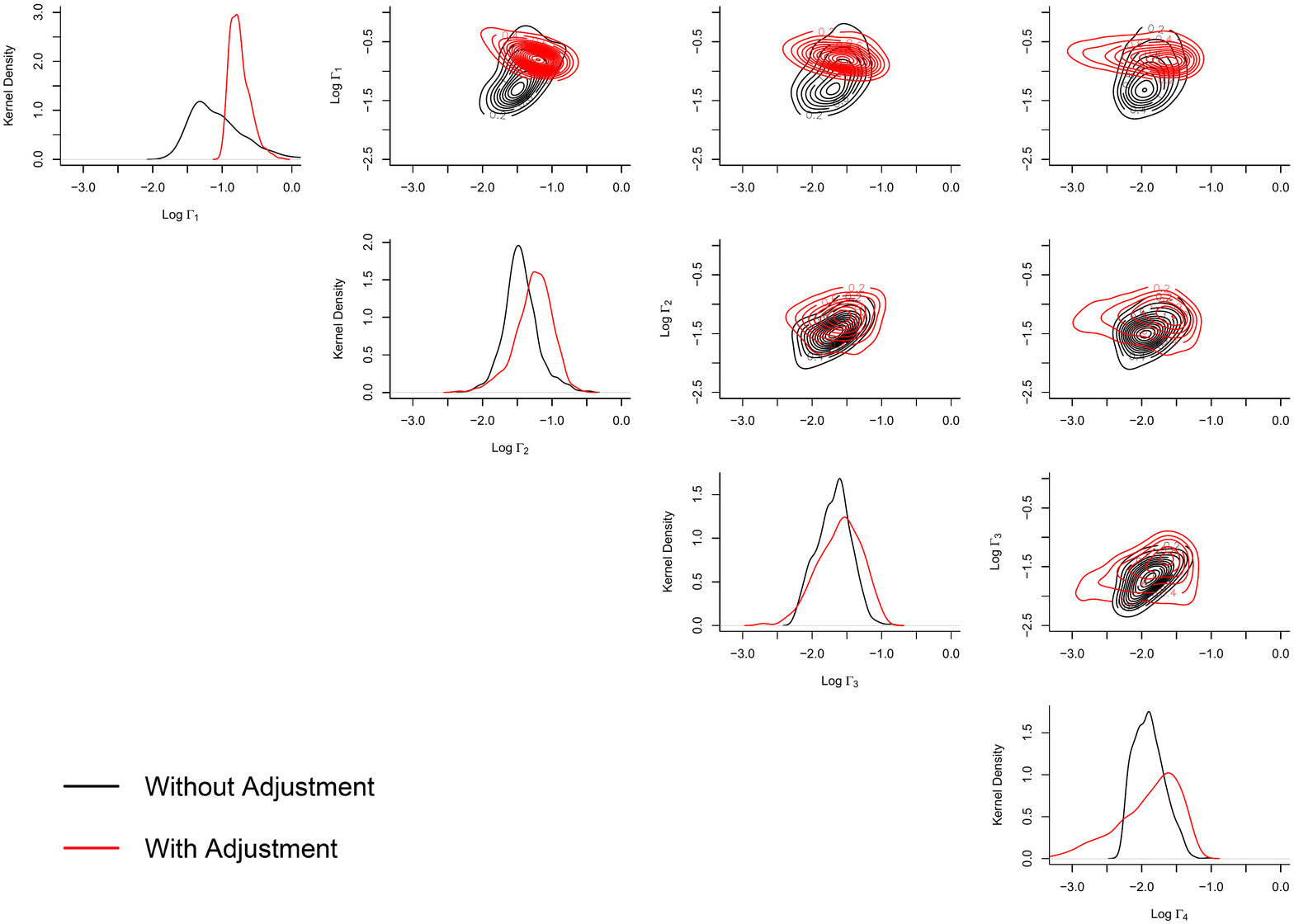} 
\end{tabular}
  \caption{\label{fig5} Univariate (diagonal panels) and bivariate (off-diagonal panels) estimated marginal posterior densities with (red) and
  without (black) total variance adjustment for $\log \Gamma_j$, $j=1,\dots, 4$.}
\end{figure}

\subsection{Approximate Gaussian process regression predictive inference using a deep neural network surrogate}\label{subsec:DNN}

Our final example considers application of a deep neural network regression model as a surrogate for Gaussian process regression
predictive inference.  We consider a $D=2$ dimensional input space.  
Suppose data $(\zvec_{\text{obs}},\Xmat_{\text{obs}})$ are observed, where $\zvec_{\text{obs}}=(z_{\text{obs},1},\dots, z_{\text{obs},n})^\top$ 
is a vector of $n$ observed values for a response $\zvec$
and $\Xmat_{\text{obs}}$ is a corresponding matrix of inputs with $n$ rows.  The $i$th row of $\Xmat_{\text{obs}}$ is 
denoted $\xvec_{\text{obs},i}$, and gives the value of the input
$\xvec=(x_1,x_2)^\top$ for the response $\zvec_{\text{obs},i}$.  We wish to make
predictive inference for a future response $\zvec^*$ for input $\xvec^*$, and we will
treat $\xvec$ as random for the purpose of deriving appropriate adjustments.  The main purpose of this example
is to show how our adjustment method applies in a setting which involves predictive inference.

For predictive inference adjustments following the framework of Section 3.2, we will start with
$R$ simulated datasets from the prior, with each dataset containing training and test samples respectively, 
\begin{align}
 (z_i^{(r)},\xvec_i^{(r)}), & \; i=1,\dots, n, \qquad (z_j^{(*,r)},\xvec_j^{(*,r)}),\; j=1,\dots, m, \label{gpreg}
\end{align}
with $r=1,\dots, R$ indexing the different replicates.  
The training and test inputs are generated independently as bivariate normal, $N(\zerovec,\Sigmamat_x)$ where the
covariance matrix $\Sigmamat_x$ has upper triangular Cholesky factor $\Rmat_x$.  $\Rmat_x$ has 
non-zero elements $r_{kl}$, $k\geq l$ which we
generated independently from an $N(0,0.5)$ density here.  $\Sigmamat_x$ is the same for all the replicates. 
The responses are generated conditionally on inputs from a 
regression model, 
\begin{align*}
  z_i^{(r)} & = f^{(r)}(\xvec_i^{(r)})+\epsilon_i^{(r)},\qquad z_j^{(*,r)}=f^{(r)}(\xvec_j^{(*,r)})+\epsilon_j^{(*,r)},
\end{align*}
where $\epsilon_i^{(r)}$, $i=1,\dots, n$, $\epsilon_j^{(*,r)}$, $j=1,\dots, m$ are independent $N(0,(\sigma^2)^{(r)})$ and 
$f^{(r)}(\cdot)$ is a mean function.  The values $(\sigma^2)^{(r)}$ are simulated 
independently from a prior $p(\sigma^2)$.  

The function $f^{(r)}(\cdot)$ is simulated from
a Gaussian process prior, and we discuss this next.
Write 
\begin{align*}
 \Xmat^{(r)} & =(\xvec_1^{(r)},\dots, \xvec_n^{(r)})^\top, \qquad \Xmat^{(*,r)}=(\xvec_1^{(*,r)},\dots, \xvec_m^{(*,r)})^\top, \qquad 
\Xmat^{(+,r)}=[{\Xmat^{(r)}}^\top\,{\Xmat^{(*,r)}}^\top]^\top, \\
\Fvec^{(r)} & = (f^{(r)}(x_1^{(r)}),\dots, f^{(r)}(x_n^{(r)}))^\top, \,\Fvec^{(*,r)}=(f^{(r)}(x_1^{(*,r)}),\dots, f^{(r)}(x_m^{(*,r)}))^\top, \,\Fvec^{(+,r)}=({\Fvec^{(r)}}^\top,{\Fvec^{(*,r)}}^\top)^\top \\
\zvec^{(r)} & = (z_1^{(r)},\dots, z_n^{(r)})^\top, \qquad \zvec^{(*,r)}=(z_1^{(*,r)},\dots, z_m^{(*,r)})^\top, \qquad
 \zvec^{(+,r)}=({\zvec^{(r)}}^\top,{\zvec^{(*,r)}}^\top)^\top.
\end{align*}
The required values for $f^{(r)}(\cdot)$ in (\ref{gpreg}) are generated under the Gaussian process prior
as
\begin{align*}
 \Fvec^{(+,r)}\sim N(\zerovec,\Cmat(\Xmat^{(+,r)},\Xmat^{(+,r)};\zetavec^{(r)}),
\end{align*}
where $\Cmat(\Xmat^{(+,r)},\Xmat^{(+,r)};\zetavec)$ is the $(m+n)\times (m+n)$ covariance matrix with entry $(i,j)$ being
$C(\xvec_i^{(+,r)},\xvec_j^{(+,r)};\zetavec)$, $\xvec_l^{(+,r)}$ denotes the $l$th row of $\Xmat^{(+,r)}$, and
$C(\xvec,\xvec';\zetavec)$ is the Mat\'{e}rn covariance function
\begin{align*}
 C(\xvec,\xvec';\zetavec) & = \tau^2 \frac{2^{1-\nu}}{\Gamma (\nu)}\left(\sqrt{2\nu}\frac{\|\xvec-\xvec '\|}{\lambda}\right)^{\nu}
  K_\nu\left(\sqrt{2\nu} \frac{\|\xvec-\xvec '\|}{\lambda}\right), 
\end{align*}
where $\|\xvec-\xvec '\|$ is the Euclidean distance between $\xvec$ and $\xvec '$, $\zetavec=(\tau^2,\lambda)$, 
$K_\nu(\cdot)$ denotes the modified Bessel function of the second kind, and $\nu$ will
be fixed at $1.5$.  For more background on Gaussian process regression see \citet{rasmussen2006gaussian}
or \citet{cressie+w11}. 

Each $\zetavec^{(r)}$, $r=1,\dots, R$, 
is simulated independently from a prior distribution in which $\tau^2$ and $\lambda$
are independent, $\tau^2\sim IG(a_\tau,b_\tau)$, $\lambda\sim IG(a_\lambda,b_\lambda)$.  Our prior
for $\sigma^2$ is inverse gamma, $IG(a_\sigma,b_\sigma)$.  Prior hyperparameters are 
$a_\sigma=3$, $b_\sigma=0.2$, $a_\tau=14.5$, $b_\tau=6.75$, $a_\lambda=b_\lambda=9$.  
This gives $E(\sigma^2)=0.1$, $\text{Var}(\sigma^2)=0.01$, $E(\tau^2)=0.1$, $\text{Var}(\tau^2)=0.02$ and 
assigns a fairly diffuse prior to the correlation between $f^{(r)}$
values at two inputs separated by unit Euclidean distance.
We also simulate one further replicate 
for which the training set values take the role of the observed data $(\zvec_{\text{obs}},\Xmat_{\text{obs}})$.   

We approximate
Gaussian process predictive inference using dense layer feed-forward neural network (FNN) regression models.  
This can be of interest since exact Gaussian process computations can be difficult for large $n$. There is a large literature
on how to do Gaussian process computations efficiently for large datasets, and we make no effort
to compare the adjustment described here with other methods for this task.  The purpose of our example here is
to illustrate the use of our adjustment method for predictive inference, where a surrogate model
is being used which is more tractable than the original one.   We consider an example with $n=1000$, 
since we wish to compare with exact Gaussian process answers.  
The neural network architecture we use has two hidden layers of size 64 each with ReLU activation functions followed 
by a univariate linear dense layer for the output. To avoid overfitting we employ $L2$ regularization and a dropout 
rate of 0.3 at the hidden layers. The unadjusted deep learning regression model is estimated using the 
R package \texttt{keras} \citep{chollet+a18} and is denoted as unadjusted DNN below when discussing the results.

\subsubsection{Adjustment and choice of conditioning set}

Consider predictive inference on a yet to be observed response value $z^*$ at a known input value $\xvec^*$
based on the observed training data.  In the notation of Section 3.2, we apply our adjustment to inference for
$\theta=z^*$.  We will use a Gaussian plug-in predictive density from the neural network at $\xvec^*$ as the
posterior approximation for $z^*$ rather than some particle approximation.  In this experiment 
one of the replicate datasets denoted $\yvec^{(i)}$ in Section 3.2 consists of a training dataset of responses and inputs of size $n$,
$(\zvec,\Xmat)$ say, as well as a test input $\xvec^*$.  
The replicate datasets $\yvec^{(i)}$ retained by the conditioning set for our adjustment will take the form $(\zvec^{(r)},\Xmat^{(r)}),\xvec_l^{(*,r)}$ for
some combinations of $r$ and $l$.  A total of $I$ replicates will be retained and we describe how the corresponding $(r,l)$ pairs are obtained below.  
The replicates $(\theta^{(i)},\yvec^{(i)})$ used in the adjustment process will be dependent here if $m>1$, because 
some of the training sets are shared between different $(\theta^{(i)},\yvec^{(i)})$ replicates, as we simulate more than one 
test input per Gaussian process replicate.  This does not complicate the estimation
of means and covariances for our adjustment, but the construction in Section 3.1
of plots for assessment using the bootstrap would need to be modified to account for the dependence.  
This will not be considered further here, since we focus on predictive inference adjustments.

To describe the conditioning set used for our adjustment method, write 
${\hat{\tau}}^{(r)}$, ${\hat{\lambda}}^{(r)}$ and ${\hat{\sigma}}^{(r)}$ for estimates of 
$\tau$, $\lambda$ and $\sigma$ for replicate $r$, $r=1,\dots, R$.  The corresponding
values for the observed data are written simply as $\hat{\tau}$, $\hat{\lambda}$, $\hat{\sigma}$.  
Estimates  of ${\tau^2}$, ${\lambda}$ are obtained based on weighted least squares fits to empirical variograms using
the  R package \texttt{geoR} \citep{ribeiro+d01}, while the estimate for $\sigma^2$ is simply the residual variance in the training data.  
Write $\mathcal{S}=\mathcal{S}(\epsilon)$ for the set of indices of those replicates $r\in \{1,\dots, R\}$ for which
\begin{align*} 
 d((\zvec,\Xmat),(\zvec_{\text{obs}},\Xmat_{\text{obs}})) &= w_1^{-2}(\hat{\tau}-{\hat{\tau}}^{(r)})^2+w_2^{-2}(\hat{\lambda}-\hat{\lambda}^{(r)})^2+w_3^{-2}(\hat{\sigma}-{\hat{\sigma}}^{(r)})^2<\epsilon,
\end{align*}
where $\epsilon>0$ is a tolerance and the weights $w_1$, $w_2$, $w_3$ are the prior standard deviations of the estimates for $\tau$, $\lambda$ 
and $\sigma$ respectively across the $R$ Gaussian process replicates.  We will choose $\epsilon$ so that $\mathcal{S}$ contains $S<R$ elements.  Then 
$\mathcal{S}$ is the index set of the $S$ replicates with covariance hyperparameter
estimates closest to the observed ones in a weighted Euclidean distance.  

Next, consider the distances 
\begin{align*}
  d_j^{(*,r)}=\min_{i=1,\dots, n}\sqrt{(\xvec_i^{(r)}-\xvec_j^{(*,r)})^\top \Sigmamat_x^{-1} (\xvec_i^{(r)}-\xvec_j^{(*,r)})},
\end{align*}
$j=1,\dots, m$, $r=1,\dots, I$.  The value $d_j^{(*,r)}$ 
is the distance of $\xvec_j^{(*,r)}$ to the nearest training input in replicate $r$.  We consider
dividing these distances for the test inputs into a set of $K$ bins.  Let
\begin{align*}
\bar{d}^r & =\frac{\max_{j,r}(d_j^{(\ast,r)})-\min_{j,r}(d_j^{(\ast,r)})}{K-1},
\end{align*} 
and define 
the bins as $B_k=[c_k,d_k)$ for $k=0,\dots, K-1$
where
\begin{align*}
 c_k & = \frac{1}{R}\sum_{r=1}^R(\min_{j,r}(d_j^{(\ast,r)})+\bar{d}^r k), \quad \frac{1}{R}\sum_{r=1}^R(\min_{j,r}(d_j^{(\ast,r)})+\bar{d}^r(k+1)).
\end{align*}
For any test input $\xvec^*$ and training set of inputs $\Xmat$ write $k(\xvec^*,\Xmat)$ for the index of the bin in which $\xvec^*$ lies for training inputs $\Xmat$.  
For a simulated training set $(\zvec,\Xmat)$ from the prior and simulated test input $\xvec '$, the conditioning set for our adjustment for predictive inference at $\xvec^*$ based on the 
observed training set is
\begin{align*}
  F(\xvec^*,\Xmat) & = \{(\zvec,\Xmat),\xvec ': d((\zvec,\Xmat),(\zvec_{\text{obs}},\Xmat_{\text{obs}}))<\epsilon, k(\xvec',\Xmat)=k(\xvec^*,\Xmat_{\text{obs}}) \}
\end{align*}
As discussed above, $\epsilon$ is chosen so that $\mathcal{S}$ has $S<R$ elements.  

With input $\xvec^*$ for the observed training data, the above conditioning set results in using
for the predictive adjustment the replicates $\yvec^{(i)}$ given by 
\begin{align*}
  & \{(\zvec^{(r)},\Xmat^{(r)}),\xvec_l^{(*,r)}: r\in \mathcal{S}, 1\leq l \leq m, k(\xvec^*,\Xmat_{\text{obs}})=k(\xvec_l^{(*,r)},\Xmat^{(r)})\}
\end{align*}
Note that the replicates used in the adjustment, and the quantities $\muvec^L$, $\muvec^R$, $\Sigmamat^L$ and $\Sigmamat_R$ required for the adjustment,
are the same for test inputs in the same bin for the observed data.  So we compute adjustments that are bin specific, applying
to all inputs in the same bin. 
The full adjustment algorithm is given in Algorithm~\ref{ex2}.  
We used $R=1000$ prior replicates, $K=10$ bins, $n=1,000$ training samples per dataset, 
$m=100$ test samples per dataset, and $S=100$.  

\subsubsection{Benchmarking and results}
\begin{figure}
\centering
\begin{tabular}{c}
    \includegraphics[width=0.9\textwidth]{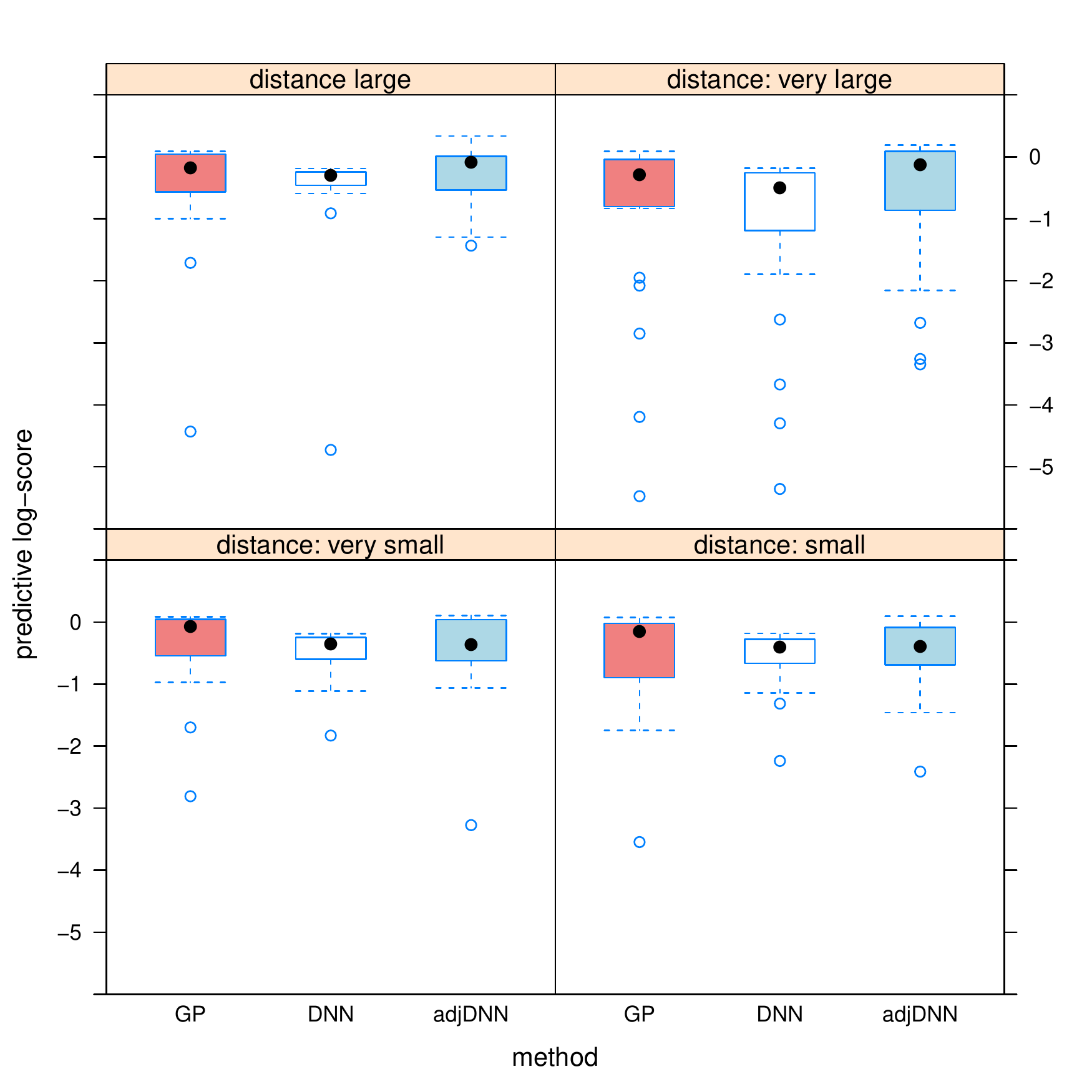} 
\end{tabular}
  \caption{\label{fig6}Boxplots of logarithmic score values across test responses 
  by the exact GP, unadjusted DNN and DNN with adjustment for test
  inputs in different bins for the observed data.  
Different panels correspond to four distance bins based on distance from the nearest training input.
The bin boundaries are equally spaced covering the range of the observed values.}
\end{figure}

To evaluate the performance of our adjustment, we compare the unadjusted DNN to our adjusted DNN inferences. 
Exact Gaussian process predictive inferences are also considered using the assumed
Mat\'{e}rn covariance function fitted using the R package \texttt{GPfit}~\citep{GPfit}.  
Figure~\ref{fig6} shows boxplots of the unadjusted, adjusted and full GP inferences for the observed data test inputs 
where performance is evaluated according to a logarithmic scoring rule.  For displaying the results, the distances of test set inputs
to the nearest training input were
partitioned into four distance intervals of equal length.  The 
adjustment procedure improves predictive inference from the neural network surrogate model, 
across all distances and in particular for points with inputs very far from the nearest training input.

\begin{algorithm}
\caption{Algorithm I, part 1}\label{ex2}
Given $R, D, n, m, K, a_{\sigma},b_{\sigma},a_{\tau},b_{\tau},a_{\lambda},b_{\lambda},\muvec_x,\Sigmamat_x$:
 \begin{algorithmic}[1]
\For{$r=1,\ldots,R$}\Comment{Loop over replicates to get bins}
\State Simulate $X^{(+,r)}$ with each row drawn independently as $N(\muvec_x,\Sigmamat_x)$.
\For{$j=1,\ldots,m$}
\State $d_j^{(\ast,r)}=\min_{i=1,\ldots,n}\sqrt{(\xvec_i^{(r)}-\xvec_j^{(\ast,r)})^T\Sigmamat_x^{-1}(\xvec_i^{(r)}-\xvec_j^{(\ast,r)})}$.
\EndFor
\EndFor
\State Define $\bar{d}^r=\frac{\max_{j,r}(d_j^{(\ast,r)})-\min_{j,r}(d_j^{(\ast,r)})}{K-1}$
\State Define  $B_k=(c_k,d_k)$, where for $k=0,\ldots,K-1$
$$c_k=\frac{1}{R}\sum_{r=1}^R((\min_{j,r}(d_j^{(\ast,r)})+\bar{d}^r k), \quad \frac{1}{R}\sum_{r=1}^R(\min_{j,r}(d_j^{(\ast,r)})+\bar{d}^r(k+1)).$$
\For{$r=1,\ldots,R$}\Comment{Loop over replicates}
\State Draw $(\sigma^2)^{(r)}\sim IG(a_{\sigma},b_{\sigma})$.
\State Draw $(\tau^2)^{(r)}\sim IG(a_{\tau},b_{\tau})$.
\State Draw $\lambda^{(r)}\sim IG(a_{\lambda},b_{\lambda})$.
\State Draw $\zvec^{(+,r)}\sim N(\zerovec,\Cmat(\Xmat^{(+,r)},\Xmat^{(+,r)};\zetavec^{(r)})+(\sigma^2)^{(r)}\Imat).$
\State Train FNN with data $\zvec^{(r)}$ and $\Xmat^{(r)}$ and construct a normal plug-in predictive density $N(\hat z^{(\ast,r)},(\hat{\sigma^2})^{(r)})$ for input $\xvec^*$, where $(\hat{\sigma^2})^{(r)}$ is the residual variance.
\EndFor
\algstore{bkbreak}
\end{algorithmic}
\end{algorithm}

\begin{algorithm}[H]
\caption{Algorithm I, part 2}
\begin{algorithmic}[1]
\algrestore{bkbreak}
\For{$k=1,\ldots,K$}\Comment{Bin specific adjustments}
\State Compute $\mu_k^L,\mu_k^R$, $\Sigma_k^L,\Sigma_k^R$ via
\[
\mu_{k,r}^L=\frac{1}{m_k^{(r)}}\sum_{l=1}^{m}z_l^{(\ast,r)}\mathbf{1}(d_{l}^{(\ast,r)}\in B_k),\quad\mu_{k,r}^R=\frac{1}{m_k^{(r)}}\sum_{l=1}^{m}\hat z_l^{(\ast,r)}\mathbf{1}(d_{l}^{(\ast,r)}\in B_k),
\]
\[
\mu_{k}^L=\frac{1}{S}\sum_{r\in \mathcal{S}}\mu_{k,r}^L,\quad\mu_{k}^R=\frac{1}{S}\sum_{r\in \mathcal{S}}\mu_{k,r}^R
\]
\[
\Sigma_{k,r}^L=\frac{1}{m_k^{(r)}-1}\sum_{l=1}^{m}((z_l^{(\ast,r)}-\mu_{k,r}^L)\mathbf{1}(d_{l}^{(\ast,r)}\in B_k))^2,
\]
\[
\Sigma_{k,r}^R=\frac{1}{m_k^{(r)}-1}\sum_{l=1}^{m}((\hat z_l^{(\ast,r)}-\mu_{k,r}^R)\mathbf{1}(d_{l}^{(\ast,r)}\in B_k))^2 + \frac{1}{m_k^{(r)}}\sum_{l=1}^{m}(\hat\sigma_{k}^2)^{(\ast,r)}\mathbf{1}(d_{l}^{(\ast,r)}\in B_k)
\]
\[
\Sigma_k^L=\frac{1}{S}\sum_{r\in \mathcal{S}}\Sigma_{k,r}^L,\quad \Sigma_k^R=\frac{1}{S}\sum_{r\in S}\Sigma_{k,r}^R
\]
where $m_k^{(r)}$ is the number of training samples in bin $B_k$ for replicate $r$.
\EndFor
\State Compute $\Sigma_k^R=C_kC_k^T$, $\Sigma_k^L-\Sigma_{k,2}^R=K_kK_k^T$, where $$\Sigma_{k,2}^R=\frac{1}{S}\sum_{r\in \mathcal{S}}\frac{1}{m_k^{(r)}}\sum_{l=1}^{m}(\hat\sigma_{k}^2)^{(\ast,l)}\mathbf{1}(d_{l}^{(\ast,r)}\in B_k).$$ 
\State \Comment{Apply adjustments to observed data}
\State Train FNN with $\zvec_{\mbox{\tiny{obs}}}$ and $\Xmat_{\text{obs}}^*$ to obtain
the plug-in predictive $N(\hat z_{\mbox{\tiny{obs}}}^{(\ast)},\hat{\sigma}^2)$ at input $\xvec^*$, where $\hat{\sigma}^2$ is the estimated residual variance for the observed data.
\State Adjust the parameters of the plug-in Gaussian predictive
\[
\hat z_{\mbox{\tiny{obs}},\mbox{\tiny{adj}}}^{(\ast)} = \mu_k^L-(\hat{z}_{\mbox{\tiny{obs}}}^{(*)}-\mu_k^R), \quad \hat\sigma_{\mbox{\tiny{adj}}}^2 = K_kC_k^{-1}  \hat\sigma^2 \mbox{ iff } d_{\text{obs}}^{\ast}\in B_k,
\]
where $d_{\text{obs}}^{\ast}=\min_{i=1,\dots, n}\sqrt{(\xvec_{\text{obs},i}-\xvec^{*})^\top \Sigmamat_x^{-1}(\xvec_{\text{obs},i}-\xvec^{*})}$.
\end{algorithmic}
\end{algorithm}

\section{Discussion}\label{sec:discussion}

We have discussed a new approach to checking and adjustment of approximate inference algorithms.  
The approach is based on using the tower property of conditional
expectation and law of total variance to relate prior and posterior means and covariances, and assessing any departure from equality when posterior means and covariances are approximated.  Adjustments can be made so that the correct relationships hold
after adjustment.  The main strength of the approach is that it can naturally handle multivariate quantities and directly
checks calibration of mean and scale parameters which are of most direct interest to users.  The main weakness is that it is based on adjusting only first and second order moments of an approximation, and is unable to correct errors in 
more complex features relating to the shape of the posterior density.

It might be possible to apply our correction first, and then use marginal quantile-based adjustments afterwards, such as those of \cite{rodrigues+ps17}, to make corrections
to the shape of marginal posterior densities.  However, this does involve additional computation.  
Higher order corrections based on the law of total cumulance \citep{brillinger69} can also be considered, 
but the lack of any general family of transformations
of a set of particles to achieve given means, covariances and higher order moments simultaneously makes it unclear how 
to generalize our approach in this direction.

\section*{Acknowledgements}

The authors thank Guilherme Rodrigues and Scott Sisson for sharing their code for the sum of log-normals example.  
David Nott was supported by a Singapore Ministry of Education Academic Research Fund Tier 1 grant (R-155-000-189-114).
Nadja Klein acknowledges support through the Emmy Noether grant KL 3037/1-1 of the German research foundation (DFG).

\bibliographystyle{apalike}
\addcontentsline{toc}{section}{\refname}\bibliography{ref} 

\end{document}